# Revisiting the Charging–Capacitor Problem: Maxwell's Equations and Approximate Solutions

**Costas J. Papachristou**

Department of Physical Sciences, Hellenic Naval Academy, Piraeus, Greece

**Abstract.** The charging capacitor is used as a standard paradigm for illustrating the concept of the Maxwell "displacement current". A certain aspect of the problem, however, is often overlooked. It concerns the conditions for satisfaction of the Faraday-Henry law both in the interior and in the exterior of the capacitor. In this article the situation is analyzed and a recursive process is described for obtaining (at least approximate) solutions of Maxwell's equations inside and outside the capacitor.

**Keywords:** Charging capacitor, Maxwell equations, Faraday-Henry law

## Introduction

The charging capacitor is used as a standard paradigm for demonstrating the significance of the Maxwell "displacement current" (see, e.g., Griffiths, 2013; Wangsness, 1986; Shadowitz, 1975; Rojansky, 1979; Jackson, 1999; McDonald, 2017; Selvan, 2009). The point is correctly made that, without this "current" term the static Ampère's law would be incomplete with regard to explaining the conservation of charge as well as the existence of electromagnetic radiation. Furthermore, the line integral of the magnetic field around a closed curve would be an ill-defined concept (see Appendix II).

A certain aspect of the problem, however, is often overlooked in the educational literature (Papachristou, 2018; Papachristou and Magoulas, 2018). It concerns the satisfaction of the Faraday-Henry law both inside and outside the capacitor. Indeed, although care is taken to ensure that the expressions used for the electromagnetic (e/m) field satisfy the Ampère-Maxwell law, no such care is exercised with regard to the Faraday-Henry law. As it turns out, the usual formulas for the e/m field satisfy this latter law only in the special case where the capacitor is being charged at a constant rate. But, if the current responsible for charging the capacitor is time-dependent, this will also be the case with the magnetic field outside the capacitor. This, in turn, implies the existence of an "induced" electric field in that region, contrary to the usual assertion that the electric field outside the capacitor is zero. Moreover, the time dependence of the magnetic field inside the capacitor is not compatible with the assumption that the electric field in that region is uniform, as the case would be in a static situation. Thus, the expressions usually given in the literature for the e/m field inside and outside a charging capacitor fail to satisfy the Faraday-Henry law in the case of a time-dependent current.

In this article we describe a method for finding expressions for the e/m field that properly satisfy the full set of Maxwell's equations (including, of course, the Faraday-Henry law) both inside and outside the capacitor. These solutions depend on two scalar functions of space and time, which functions satisfy a certain system of partial differential equations (PDEs). The time-dependent current that charges the capacitor appears as a sort of parametric function in this system.

We suggest a mathematical process for obtaining solutions of the above-mentioned system of PDEs in the form of power series with respect to time. This allows one to find approximate expressions for the e/m field in certain situations. For example, a slowly varying (thus almost time-independent) current allows for the "classical" (albeit incorrect in precise terms) solutions given in the literature, while a current that is almost linearly dependent on time (as may be assumed, in general, for any smoothly varying current in a very short time period) allows for new solutions that correct the standard expressions for the electric field while retaining the corresponding expressions for the magnetic field.

It should be noted that, regarding the solutions in the exterior of the capacitor, no retardation effects related to the finite speed of propagation of e/m interactions will concern us here. As discussed at the end of the article, our solutions are valid at points of space not far from the capacitor, so that any change in the physical system will practically be felt simultaneously at all points of interest.

## Solutions of Maxwell's equations inside the capacitor

We consider a parallel-plate capacitor with circular plates of radius $a$, thus of area $A=\pi a^2$. The space in between the plates is assumed to be empty of matter. The capacitor is being charged by a time-dependent current $I(t)$ flowing in the $+z$ direction (see Fig. 1). The $z$-axis is perpendicular to the plates (the latter are therefore parallel to the $xy$-plane) and passes through their centers, as seen in the figure (by $\hat{u}_z$ we denote the unit vector in the $+z$ direction).

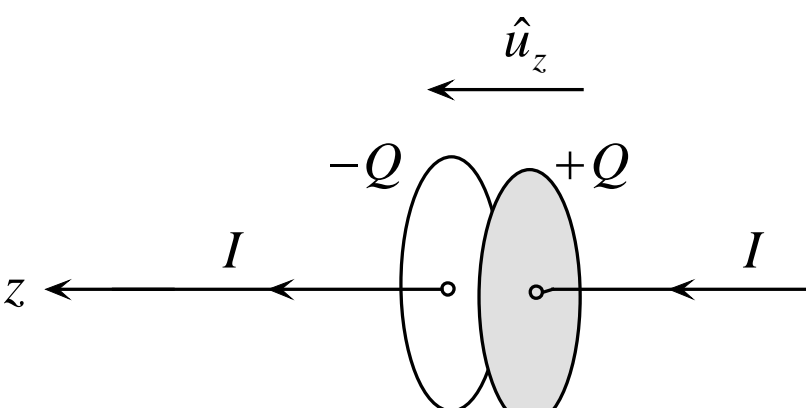

**Figure 1.** A parallel-plate capacitor charged by a current $I(t)$.

The capacitor is being charged at a rate $dQ/dt=I(t)$, where $+Q(t)$ is the charge on the right plate (as seen in the figure) at time $t$. If $\sigma(t)=Q(t)/\pi a^2=Q(t)/A$ is the surface charge density on the right plate, then the time derivative of $\sigma$ is given by

$$\sigma'(t)=\frac{Q'(t)}{A}=\frac{I(t)}{A} \tag{1}$$

We assume that the plate separation is very small compared to the radius $a$, so that the e/m field inside the capacitor is practically independent of $z$, although it *does* depend on the normal distance $\rho$ from the $z$-axis. In cylindrical coordinates $(\rho,\varphi,z)$ the magnitude of the e/m field at any time $t$ will thus only depend on $\rho$ (due to the symmetry of the problem, this magnitude will not depend on the angle $\varphi$).

We assume that the positive and the negative plate of the capacitor of Fig. 1 are centered at $z=0$ and $z=d$, respectively, on the $z$-axis, where, as mentioned above, the plate separation $d$ is much smaller than the radius $a$ of the plates. The interior of the capacitor is then the region of space with $0\le\rho<a$ and $0<z<d$.

The magnetic field inside the capacitor is azimuthal, of the form $\vec{B} = B(\rho,t)\,\hat{u}_\varphi$. A standard practice in the literature is to assume that, at all *t*, the electric field in this region is uniform, of the form

$$\vec{E} = \frac{\sigma(t)}{\varepsilon_0}\,\hat{u}_z \tag{2}$$

while everywhere outside the capacitor the electric field vanishes. With this assumption the magnetic field inside the capacitor is found to be (Wangsness, 1986; Shadowitz, 1975; McDonald, 2017)

$$\vec{B} = \frac{\mu_0 I(t)\rho}{2\pi a^2}\,\hat{u}_\varphi = \frac{\mu_0 I(t)\rho}{2A}\,\hat{u}_\varphi \tag{3}$$

Expressions (2) and (3) must, of course, satisfy the Maxwell system of equations in empty space, which system we choose to write in the form (Griffiths, 2013; Papachristou, 2020)

$$\begin{aligned} &(a)\quad \vec{\nabla}\cdot\vec{E} = 0 &\qquad &(c)\quad \vec{\nabla}\times\vec{E} = -\frac{\partial \vec{B}}{\partial t} \\ &(b)\quad \vec{\nabla}\cdot\vec{B} = 0 &\qquad &(d)\quad \vec{\nabla}\times\vec{B} = \varepsilon_0\mu_0\,\frac{\partial \vec{E}}{\partial t} \end{aligned} \tag{4}$$

By using cylindrical coordinates (see Appendix I) and by taking (1) into account, one may show that (2) and (3) satisfy three of Eqs. (4), namely, (*a*), (*b*) and (*d*). This is not the case with the Faraday-Henry law (4*c*), however, since by (2) and (3) we find that $\vec{\nabla}\times\vec{E} = 0$ while

$$\frac{\partial \vec{B}}{\partial t} = \frac{\mu_0 I'(t)\rho}{2A}\,\hat{u}_\varphi \;.$$

An exception occurs if the current *I* is constant in time, i.e., if the capacitor is being charged at a constant rate, so that $I'(t)=0$. This is actually the assumption silently or explicitly made in many textbooks [see, e.g., Wangsness (1986), Chap. 21]. But, for a current *I*(*t*) with arbitrary time dependence, the pair of fields (2) and (3) does not satisfy the third Maxwell equation.

To remedy the situation and restore the validity of the full set of Maxwell's equations in the interior of the capacitor, we must somehow correct the above expressions for the e/m field. To this end we employ the following *Ansatz*, taking into account Lemma 1 in Appendix III:

$$\vec{E} = \left( \frac{\sigma(t)}{\varepsilon_0} + f(\rho,t) \right) \hat{u}_z \ ,$$
$$\vec{B} = \left( \frac{\mu_0 I(t) \rho}{2A} + g(\rho,t) \right) \hat{u}_\varphi \ ; \qquad (5)$$
$$\sigma'(t) = I(t)/A$$

where $f(\rho,t)$ and $g(\rho,t)$ are functions to be determined consistently with the given current function $I(t)$ and the given initial conditions. It can be checked that the solutions (5) automatically satisfy the first two Maxwell equations (4*a*) and (4*b*). By the Faraday-Henry law (4*c*) and the Ampère-Maxwell law (4*d*) we get the following system of PDEs:

$$\frac{\partial f}{\partial \rho} = \frac{\partial g}{\partial t} + \frac{\mu_0 I'(t) \rho}{2A}$$
$$\frac{1}{\rho}\frac{\partial(\rho g)}{\partial \rho} = \varepsilon_0 \mu_0 \frac{\partial f}{\partial t} \qquad (6)$$

Note in particular that the "classical" solution with $f(\rho,t)\equiv 0$ and $g(\rho,t)\equiv 0$ is possible only if $I'(t)=0$, i.e., if the current $I$ is constant in time, which means that the capacitor is being charged at a constant rate.

The quantity $(1/\rho)\partial(\rho g)/\partial\rho$ in the second equation, having its origin at the expression for $\vec{\nabla} \times \vec{B}$ in cylindrical coordinates, must tend to a finite limit for $\rho \to 0$ in order that the *rot* of the magnetic field be finite at the center of the capacitor. For this to be the case, $\partial(\rho g)/\partial\rho$ must only contain terms of at least first order in $\rho$. This, in turn, requires that $g$ itself must be of at least first order (i.e., linear with no constant term) in $\rho$ for all $t$, or else $g$ must be identically zero. We must, therefore, require that

$$g(\rho,t) \to 0 \ \text{ for } \ \rho \to 0 \qquad (7)$$

for all $t$. Keeping this condition in mind, we can rewrite the system (6) in a more symmetric form:

$$\frac{\partial f}{\partial \rho} = \frac{\partial g}{\partial t} + \frac{\mu_0 I'(t) \rho}{2A}$$
$$\frac{\partial(\rho g)}{\partial \rho} = \varepsilon_0 \mu_0 \frac{\partial(\rho f)}{\partial t} \qquad (8)$$

In principle, one needs to solve the system (8) for a given current $I(t)$ and for given initial conditions. An alternative approach, leading to approximate solutions of various forms, is to expand all functions (i.e., $f$, $g$ and $I$) in powers of time, $t$. We thus write:

$$I(t) = \sum_{n=0}^{\infty} I_n t^n \qquad (9a)$$

$$f(\rho,t)=\sum_{n=0}^{\infty} f_n(\rho)\,t^n \tag{9b}$$

$$g(\rho,t)=\sum_{n=0}^{\infty} g_n(\rho)\,t^n \tag{9c}$$

Then, for example,

$$I'(t)=\sum_{n=1}^{\infty} nI_n\,t^{n-1}=\sum_{n=0}^{\infty}(n+1)I_{n+1}\,t^n\ ,\ \text{etc.}$$

Obviously, $I_n$ has dimensions of current × (time)$^{-n}$, while $f_n$ and $g_n$ have dimensions of field intensity (electric and magnetic, respectively) × (time)$^{-n}$.

Substituting the series expansions (9) into the system (8), and equating coefficients of similar powers of $t$ on both sides of the ensuing equations, we get a recursion relation in the form of a system of PDEs:

$$\begin{aligned} f_n{}'(\rho) &= (n+1)\left[\, g_{n+1}(\rho)+\frac{\mu_0\rho}{2A}I_{n+1}\right] \\ \left[\rho g_n(\rho)\right]' &= (n+1)\,\varepsilon_0\mu_0\,\rho f_{n+1}(\rho) \end{aligned} \tag{10}$$

for $n$=0,1,2,... All non-vanishing functions $g_n(\rho)$ are required to satisfy the boundary condition (7); i.e., $g_n(\rho)\to 0$ for $\rho\to 0$.

An obvious solution of the system (10) is the trivial solution $f_n(\rho)\equiv 0$ and $g_n(\rho)\equiv 0$ for all $n$=0,1,2,..., corresponding to $f(\rho,t)\equiv 0$ and $g(\rho,t)\equiv 0$. For this to be the case, we must have $I_{n+1}=0$ for all $n$=0,1,2,..., which means that $I(t)=I_0$=constant (independent of $t$). This is the case typically treated in the literature, although the condition $I=const.$ is usually not stated explicitly.

The simplest nontrivial solution of the problem is found by assuming that $f$ and $g$ are time-independent, i.e., are functions of $\rho$ only. Then, by (9*b*) and (9*c*), $f=f_0(\rho)$ and $g=g_0(\rho)$, while $f_n(\rho)=0$ and $g_n(\rho)=0$ for $n>0$. The system (10) for $n=0$ gives

$$f_0{}'(\rho)=\frac{\mu_0 I_1\rho}{2A}\quad \text{and}\quad \left[\rho g_0(\rho)\right]'=0$$

with solutions

$$f_0(\rho)=\frac{\mu_0 I_1\rho^2}{4A}+C\quad \text{and}\quad g_0(\rho)=\frac{\lambda}{\rho}\ ,$$

respectively. The boundary condition $g_0(\rho)\to 0$ for $\rho\to 0$ cannot be satisfied for $\lambda\neq 0$; we are thus compelled to set $\lambda=0$. Given that $f(\rho,t)=f_0(\rho)$ and $g(\rho,t)=g_0(\rho)$, the solution of the system (8) is

$$f(\rho,t)=\frac{\mu_0 I_1\rho^2}{4A}+C\ ,\quad g(\rho,t)\equiv 0 \tag{11}$$

As is easy to check, by the first of Eqs. (10) it follows that $I_n$=0 for $n$>1. Therefore $I(t)$ is linear in $t$, i.e., is of the form $I(t)=I_0+I_1 t$. By assuming the initial condition $I(0)=0$, we have that $I_0$=0 and

$$I(t) = I_1\, t \tag{12}$$

On the other hand, by integrating Eq. (1): $\sigma'(t)=I(t)/A$, and by assuming that the capacitor is initially uncharged [$\sigma(0)=0$], we get:

$$\sigma(t) = \frac{I_1 t^2}{2A} \tag{13}$$

Finally, by Eqs. (5), (11), (12) and (13) the e/m field in the interior of the capacitor is

$$\vec{E} = \left( \frac{I_1 t^2}{2\varepsilon_0 A} + \frac{\mu_0 I_1 \rho^2}{4A} \right) \hat{u}_z \ , \qquad \vec{B} = \frac{\mu_0 I_1 t \rho}{2A}\, \hat{u}_\varphi \tag{14}$$

where we have set $C$=0 since, in view of the assumed initial conditions, there is no electric field inside the capacitor if $I_1$=0. In order for the solution (14) to be valid, the current $I(t)$ charging the capacitor must vary linearly with time, according to (12).

## Solutions of Maxwell's equations outside the capacitor

We recall that the positive and the negative plate of the capacitor of Fig. 1 are centered at $z$=0 and $z$=$d$, respectively, on the $z$-axis, where the plate separation $d$ is much smaller than the radius $a$ of the plates. The space exterior to the capacitor consists of points with $\rho > 0$ and $z \notin (0,d)$, as well as points with $\rho > a$ and $0 < z < d$. (In the former case we exclude points on the $z$-axis, with $\rho$=0, to ensure the finiteness of our solutions in that region.) We assume that the current $I(t)$ is of "infinite" extent and hence the magnitude of the e/m field is practically $z$-independent.

The e/m field outside the capacitor is usually described mathematically by the equations (Wangsness, 1986; Shadowitz, 1975; McDonald, 2017)

$$\vec{E} = 0 \ , \quad \vec{B} = \frac{\mu_0 I(t)}{2\pi\rho}\, \hat{u}_\varphi \tag{15}$$

As the case is with the standard solutions in the interior of the capacitor, the solutions (15) fail to satisfy the Faraday-Henry law (4$c$) (although they do satisfy the remaining three Maxwell equations), since $\vec{\nabla} \times \vec{E} = 0$ while

$$\frac{\partial \vec{B}}{\partial t} = \frac{\mu_0 I'(t)}{2\pi\rho}\, \hat{u}_\varphi \ .$$

As before, an exception occurs if the current $I$ is constant in time, i.e., if the capacitor is being charged at a constant rate, so that $I'(t)=0$.

To find more general solutions that satisfy the entire set of the Maxwell equations, we work as in the previous section. Taking into account Lemma 2 in Appendix III, we assume the following general form of the e/m field everywhere outside the capacitor:

$$\vec{E} = f(\rho,t)\,\hat{u}_z \ , \qquad \vec{B} = \left( \frac{\mu_0 I(t)}{2\pi\rho} + g(\rho,t) \right) \hat{u}_\varphi \tag{16}$$

where $f$ and $g$ are functions to be determined consistently with the given current function $I(t)$. The solutions (16) automatically satisfy the first two Maxwell equations (4*a*) and (4*b*). By Eqs. (4*c*) and (4*d*) we get the following system of PDEs:

$$\frac{\partial f}{\partial \rho} = \frac{\partial g}{\partial t} + \frac{\mu_0 I'(t)}{2\pi\rho} \ , \qquad \frac{\partial(\rho g)}{\partial \rho} = \varepsilon_0 \mu_0 \frac{\partial(\rho f)}{\partial t} \tag{17}$$

Again, the usual solution with $f(\rho,t)\equiv 0$ and $g(\rho,t)\equiv 0$ is possible only if $I'(t)=0$, i.e., if the capacitor is being charged at a constant rate. Note also that, since now $\rho\neq 0$, the boundary condition (7) for $g$ no longer applies.

As we did in the previous section, we seek a series solution of the system (17) in powers of $t$. We thus expand $f$, $g$ and $I$ as in Eqs. (9), substitute the expansions into the system (17), and compare terms with equal powers of $t$. The result is a new recursive system of PDEs:

$$f_n'(\rho) = (n+1)\left[ g_{n+1}(\rho) + \frac{\mu_0}{2\pi\rho} I_{n+1} \right] \ , \qquad \left[\rho g_n(\rho)\right]' = (n+1)\,\varepsilon_0 \mu_0 \,\rho f_{n+1}(\rho) \tag{18}$$

for $n=0,1,2,...$ Again, an obvious solution is the trivial solution $f_n(\rho)\equiv 0$ and $g_n(\rho)\equiv 0$ for all $n=0,1,2,...$, corresponding to $f(\rho,t)\equiv 0$ and $g(\rho,t)\equiv 0$. This requires that $I_{n+1}=0$ for all $n=0,1,2,...$, so that $I(t)=I_0=$constant (independent of $t$).

As we did previously, we seek time-independent solutions for $f$ and $g$, so that $f=f_0(\rho)$ and $g=g_0(\rho)$ while $f_n(\rho)=0$ and $g_n(\rho)=0$ for $n>0$. The system (18) for $n=0$ gives

$$f_0'(\rho) = \frac{\mu_0 I_1}{2\pi\rho} \quad \text{and} \quad \left[\rho g_0(\rho)\right]' = 0$$

with solutions

$$f_0(\rho) = \frac{\mu_0 I_1}{2\pi} \ln(\kappa\rho) \quad \text{and} \quad g_0(\rho) = \frac{\lambda}{2\pi\rho} \ ,$$

respectively (remember that $\rho>0$), where $\kappa$ is a positive constant quantity having dimensions of inverse length, and where a factor of $2\pi$ has been put in $g_0(\rho)$ for future convenience. Given that $f(\rho,t)=f_0(\rho)$ and $g(\rho,t)=g_0(\rho)$, the solution of the system (17) is

$$f(\rho,t)=\frac{\mu_0 I_1}{2\pi}\ln(\kappa\rho)\ , \quad g(\rho,t)=\frac{\lambda}{2\pi\rho} \tag{19}$$

By the first of Eqs. (18) it follows that $I_n=0$ for $n>1$. Therefore $I(t)$ is linear in $t$, of the form $I(t)=I_0+I_1 t$. By assuming the initial condition $I(0)=0$, we have that $I_0=0$ and

$$I(t)=I_1\, t \tag{20}$$

In view of the above results, the e/m field (16) in the exterior of the capacitor is

$$\begin{aligned}\vec{E}&=\frac{\mu_0 I_1}{2\pi}\ln(\kappa\rho)\,\hat{u}_z\ ,\\ \vec{B}&=\frac{\mu_0 I_1 t+\lambda}{2\pi\rho}\,\hat{u}_\varphi\end{aligned} \tag{21}$$

For this solution to be valid, the current $I(t)$ must vary linearly with time.

By comparing Eqs. (14) and (21) we observe that the value of the electric field inside the capacitor does not match the value of this field outside for $\rho=a$, where $a$ is the radius of the capacitor. This discontinuity of the electric field at the boundary of the space occupied by the capacitor is a typical characteristic of capacitor problems, in general. On the other hand, in order that the magnetic field in the strip $0<z<d$ be continuous for $\rho=a$, the expression for $\vec{B}$ in (21) must match the corresponding expression in (14) upon substituting $\rho=a$ and by taking into account that $A=\pi a^2$. This requires that we set $\lambda=0$ in (21), so that this equation finally becomes

$$\begin{aligned}\vec{E}&=\frac{\mu_0 I_1}{2\pi}\ln(\kappa\rho)\,\hat{u}_z\ ,\\ \vec{B}&=\frac{\mu_0 I_1 t}{2\pi\rho}\,\hat{u}_\varphi\end{aligned} \tag{22}$$

## Discussion

As we have seen, expressions for the e/m field inside and outside a charging capacitor may be sought in the general form given by Eqs. (5) and (16), respectively. These expressions contain two unknown functions $f(\rho,t)$ and $g(\rho,t)$ which, in view of Maxwell's equations, satisfy the systems of PDEs (8) and (17). These PDEs, in turn, admit series solutions in powers of $t$, of the form (9), where it is assumed that the current $I(t)$ itself may be expanded in this fashion.

The coefficients of expansion of $f$ and $g$ may be determined, in principle, by means of the recursion relations (10) and (18), both of which are of the general form

$$f_n{}'(\rho) = (n+1)\left[ g_{n+1}(\rho) + h(\rho) I_{n+1} \right]$$
$$\left[ \rho g_n(\rho) \right]' = (n+1)\, \varepsilon_0 \mu_0\, \rho f_{n+1}(\rho) \qquad (23)$$

This is not an easy system to integrate, so we are compelled to make certain *ad hoc* assumptions. Suppose, e.g., that we seek a solution such that $f_n(\rho)=0$ and $g_n(\rho)=0$ for $n>k$ ($k\geq 0$). It then follows from the first of Eqs. (23) that $I_{n+1}=0$ for $n>k$ or, equivalently, $I_n=0$ for $n>k+1$. Thus, if $k=0$, $I(t)$ must be linear in $t$; if $k=1$, $I(t)$ must be quadratic in $t$; etc.

For a current varying sufficiently slowly with time, we may approximately assume that $I_n=0$ for $n>0$, so that $I(t)=I_0=const$. This allows for the possibility that $f$ and $g$ vanish identically, as is effectively assumed (though not always stated explicitly) in the literature [see, however, Milsom (2020)]. On the other hand, any smoothly varying $I(t)$ may be assumed to vary linearly with time for a very short time period. Then, a solution of the form (14) and (22) is admissible.

There are several aspects of the solutions described by Eqs. (14) and (22) that may look unphysical: (*a*) the electric field in (22) apparently diverges for $\rho\to\infty$; (*b*) the magnetic field in both (14) and (22) diverges for $t\to\infty$; (*c*) although, by assumption, there are no charges at the interface between the interior and the exterior of the capacitor (i.e., on the cylindrical surface defined by $0<z<d$ and $\rho=a$) the electric field is non-continuous on that surface, contrary to the general boundary conditions required by Maxwell's equations; (*d*) the constant $\kappa$ in (22) appears to be arbitrary. We may thus use the above solutions only as approximate ones for values of $\rho$ not much larger than the radius $a$ of the plates, as well as for short time intervals. (Note that $\rho$ has to be much smaller than the length of the wire that charges the capacitor if this wire is to be considered of "infinite" length, hence if the external e/m field is to be regarded as $z$-independent.) We may smoothen the discontinuity problem of the electric field for $\rho=a$ by assuming that this field is continuous at $t=0$, i.e., at the moment when the charging of the capacitor begins. By setting $\rho=a$ in (14) and (22) and by equating the corresponding expressions for $\vec{E}$ we may then determine the value of the constant $\kappa$ in (22). The result is: $\kappa=e^{1/2}/a$.

For an enlightening discussion of the subtleties concerning the e/m field produced by an infinitely long straight current, the reader is referred to Example 7.9 of Griffiths (2013).

## Acknowledgment

I thank Aristidis N. Magoulas for a number of helpful discussions on this subject.

## Appendix I. Vector operators in cylindrical coordinates

Let $\vec{A}$ be a vector field, expressed in cylindrical coordinates $(\rho, \varphi, z)$ as

$$\vec{A} = A_\rho(\rho,\varphi,z)\,\hat{u}_\rho + A_\varphi(\rho,\varphi,z)\,\hat{u}_\varphi + A_z(\rho,\varphi,z)\,\hat{u}_z \ .$$

The *div* and the *rot* of this field in this system of coordinates are written, respectively, as follows:

$$\vec{\nabla}\cdot\vec{A}=\frac{1}{\rho}\frac{\partial}{\partial\rho}(\rho A_\rho)+\frac{1}{\rho}\frac{\partial A_\varphi}{\partial\varphi}+\frac{\partial A_z}{\partial z}\ ,$$

$$\vec{\nabla}\times\vec{A}=\left(\frac{1}{\rho}\frac{\partial A_z}{\partial\varphi}-\frac{\partial A_\varphi}{\partial z}\right)\hat{u}_\rho+\left(\frac{\partial A_\rho}{\partial z}-\frac{\partial A_z}{\partial\rho}\right)\hat{u}_\varphi+\frac{1}{\rho}\left(\frac{\partial}{\partial\rho}(\rho A_\varphi)-\frac{\partial A_\rho}{\partial\varphi}\right)\hat{u}_z\ .$$

In particular, if the vector field is of the form

$$\vec{A}=A_\varphi(\rho)\,\hat{u}_\varphi+A_z(\rho)\,\hat{u}_z\ ,$$

then $\vec{\nabla}\cdot\vec{A}=0$.

## Appendix II. Charging capacitor: The "textbook" approach

When writing the Ampère-Maxwell law in its integral form, one must carefully define the concept of the *total current through a loop C* (where by "loop" we mean a closed curve in space).

*Proposition.* Consider a region *R* of space within which the distribution of charge, expressed by the volume charge density, is time-independent. Let *C* be an oriented loop in *R*, and let *S* be any open surface in *R* bordered by *C* and oriented accordingly. We define the total current through *C* as the surface integral of the current density $\vec{J}$ over $S$:

$$I_{in}=\int_S\vec{J}\cdot\overrightarrow{da} \tag{A.1}$$

Then, the quantity $I_{in}$ has a well-defined value independent of the particular choice of *S* (that is, $I_{in}$ is the same for all open surfaces *S* bounded by $C$).

*Proof.* By the equation of continuity for the electric charge [see, e.g., Papachristou (2020), Chap. 6] and by the fact that the charge density inside the region *R* is static, we have that $\vec{\nabla}\cdot\vec{J}=0$. Therefore, within this region of space the current density has the properties of a solenoidal field. In particular, the value of the surface integral of $\vec{J}$ will be the same for all open surfaces *S* sharing a common border *C*.

As an example, let us consider a circuit carrying a time-dependent current *I*(*t*). If the circuit does not contain a capacitor, no charge is piling up at any point and the charge density at any elementary segment of the circuit is constant in time. Moreover, at each instant *t*, the current *I* is constant along the circuit, its value changing only with time. Now, if *C* is a loop encircling some section of the circuit, as shown in Fig. 2, then, at each instant *t*, the same current *I*(*t*) will pass through any open surface *S* bordered by *C*. Thus, the integral in (A.1) is well defined for all *t*, assuming the same value $I_{in}=I(t)$ for all *S*.

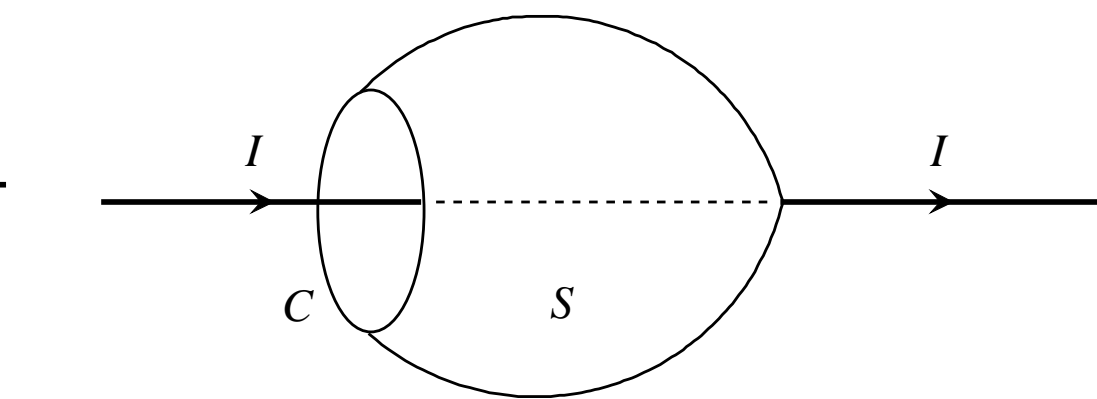

**Figure 2.** An open surface $S$ bordered by a closed curve $C$ encircling a current $I$.

Things change if the circuit contains a capacitor that is charging or discharging. It is then no longer true that the current $I(t)$ is constant along the circuit; indeed, $I(t)$ is zero inside the capacitor and nonzero outside. Thus, the value of the integral in (A.1) depends on whether the surface $S$ does or does not contain points belonging to the interior of the capacitor.

Figure 3 shows a simple circuit containing a capacitor that is being charged by a time-dependent current $I(t)$. At time $t$, the plates of the capacitor, each of area $A$, carry charges $\pm Q(t)$.

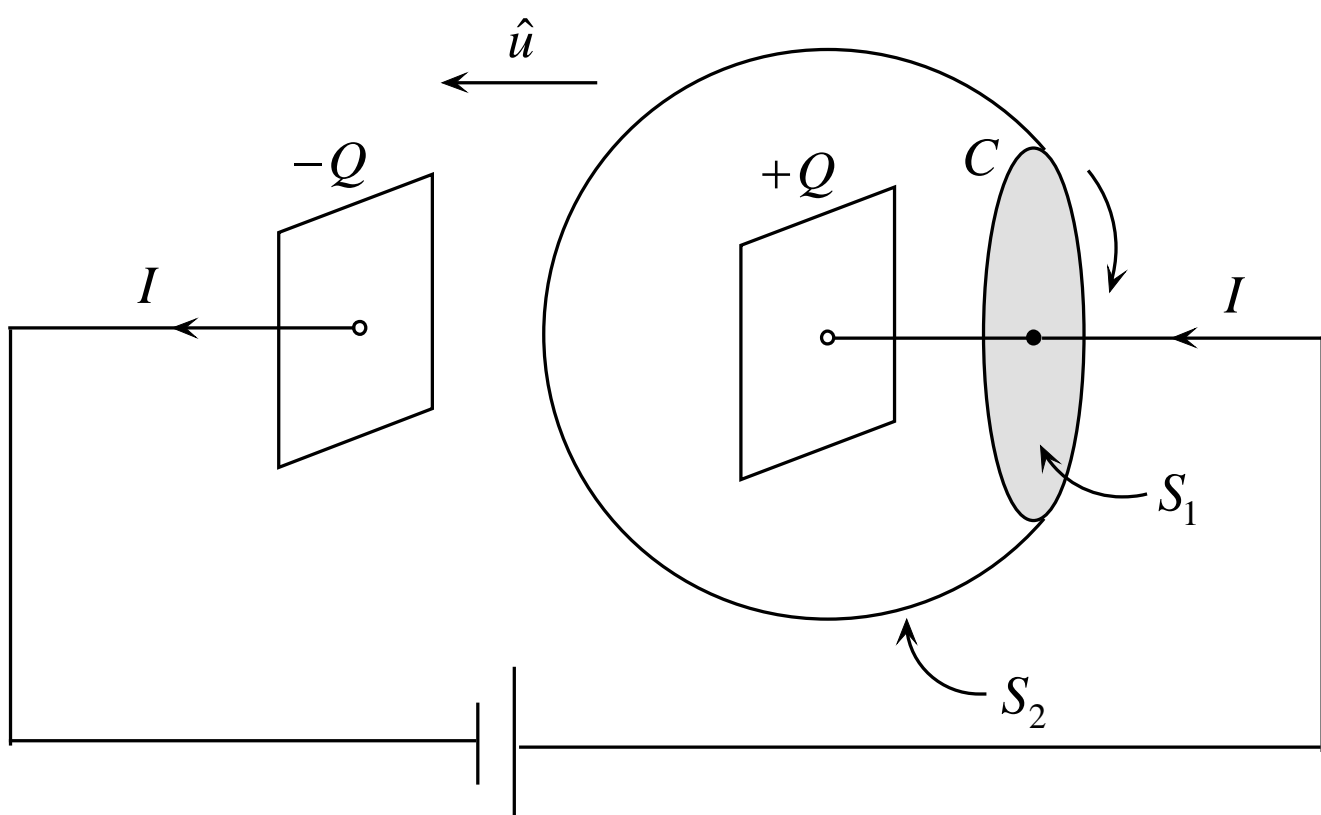

**Figure 3.** A circuit containing a capacitor charged by a current $I(t)$.

Assume that we encircle the current $I$ by an imaginary plane loop $C$ parallel to the positive plate and oriented in accordance with the "right-hand rule", consistently with the direction of $I$ (this direction is indicated by the unit vector $\hat{u}$). The "current through $C$" is here an ill-defined notion since the value of the integral in Eq. (A.1) is $I_{in}=I$ for the flat surface $S_1$ and $I_{in}=0$ for the curved surface $S_2$. This, in turn, implies that Ampère's law of magnetostatics [see, e.g., Papachristou (2020), Chap. 7] cannot be valid in this case, given that, according to this law, the integral of the magnetic field $\vec{B}$ along the loop $C$, equal to $\mu_0 I_{in}$ , would not be uniquely defined but would depend on the choice of the surface $S$ bounded by $C$.

Maxwell restored the single-valuedness of the closed line integral of $\vec{B}$ by introducing the so-called *displacement current*, which is essentially the rate of change of a time-dependent electric field:

$$\vec{J}_d = \varepsilon_0 \frac{\partial \vec{E}}{\partial t} \quad \Leftrightarrow \quad I_d = \int_S \vec{J}_d \cdot \overrightarrow{da} = \varepsilon_0 \int_S \frac{\partial \vec{E}}{\partial t} \cdot \overrightarrow{da} \qquad \text{(A.2)}$$

The *Ampère-Maxwell law* reads:

$$\vec{\nabla}\times\vec{B}=\mu_0\vec{J}+\varepsilon_0\mu_0\frac{\partial\vec{E}}{\partial t}\quad\Leftrightarrow$$

$$\oint_C \vec{B}\cdot\overrightarrow{dl}=\mu_0 I_{in}+\varepsilon_0\mu_0\int_S\frac{\partial\vec{E}}{\partial t}\cdot\overrightarrow{da}\equiv\mu_0\,(I+I_d)_{in} \tag{A.3}$$

where $I_{in}$ is given by Eq. (A.1).

Now, the standard "textbook" approach to the charging capacitor problem goes as follows: Outside the capacitor the electric field vanishes everywhere, while inside the capacitor the electric field is uniform – albeit time-dependent – and has the static-field-like form

$$\vec{E}=\frac{\sigma(t)}{\varepsilon_0}\,\hat{u}\;=\;\frac{Q(t)}{\varepsilon_0 A}\,\hat{u} \tag{A.4}$$

where $\sigma(t)=Q(t)/A$ is the surface charge density on the positive plate at time $t$. This density is related to the current $I$ that charges the capacitor by

$$\sigma'(t)=\frac{Q'(t)}{A}=\frac{I(t)}{A} \tag{A.5}$$

(the prime indicates differentiation with respect to $t$). Thus, inside the capacitor,

$$\frac{\partial\vec{E}}{\partial t}=\frac{\sigma'(t)}{\varepsilon_0}\,\hat{u}\;=\;\frac{I(t)}{\varepsilon_0 A}\,\hat{u} \tag{A.6}$$

Outside the capacitor the time derivative of the electric field vanishes everywhere and, therefore, so does the displacement current.

Now, on the flat surface $S_1$ the total current through $C$ is $(I+I_d)_{in}=I+0=I(t)$. The Ampère-Maxwell law (A.3) then yields:

$$\int_C \vec{B}\cdot\overrightarrow{dl}=\mu_0 I(t) \tag{A.7}$$

On the curved surface $S_2$ the total current through $C$ is $(I+I_d)_{in}=0+I_{d,in}=I_{d,in}$ , where the quantity on the right assumes a nonzero value only for the portion $S_2'$ of $S_2$ that lies inside the capacitor. This quantity is equal to

$$I_{d,in}=\varepsilon_0\int_{S_2'}\frac{\partial\vec{E}}{\partial t}\cdot\overrightarrow{da}=\frac{I(t)}{A}\int_{S_2'}\hat{u}\cdot\overrightarrow{da} \tag{A.8}$$

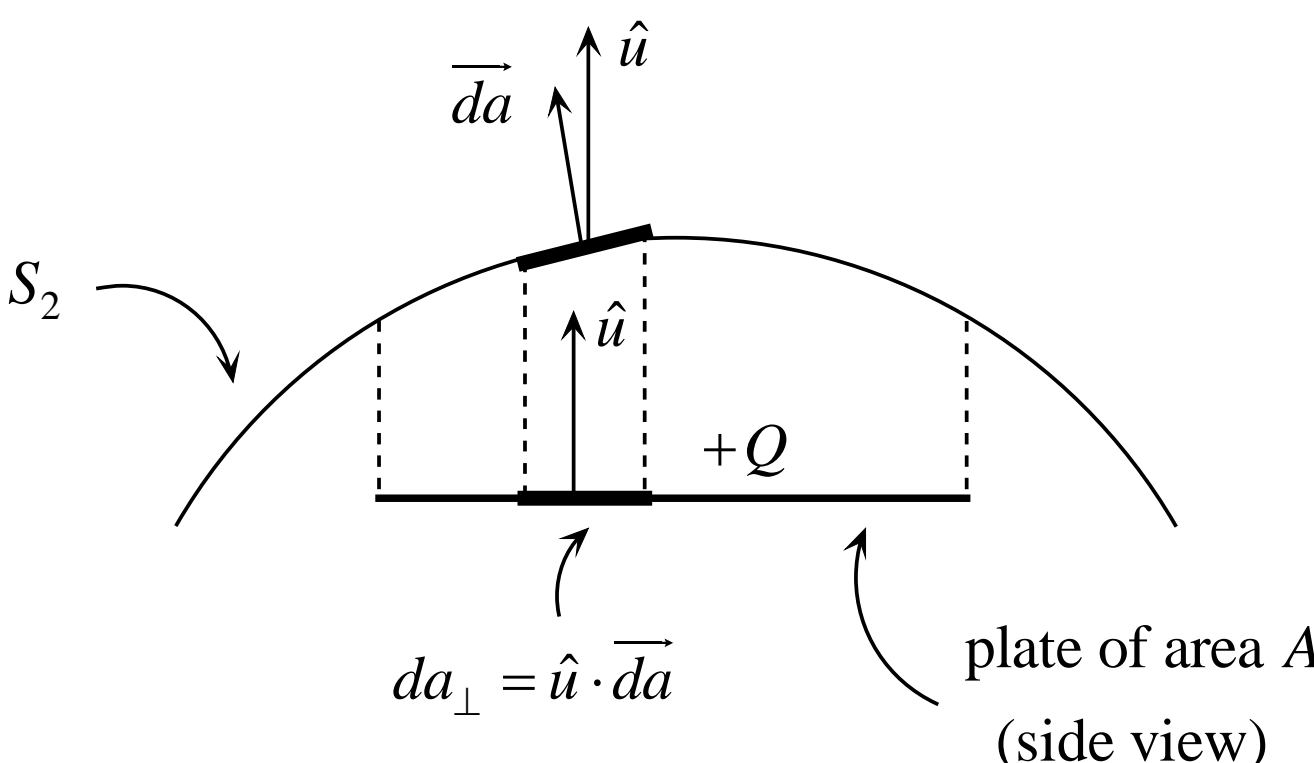

**Figure 4.** A side view of the positively-charged plate of the capacitor.

The dot product in the integral on the right of (A.8) represents the projection of the surface element $\overrightarrow{da}$ onto the axis defined by the unit vector $\hat{u}$ (see Fig. 4). This is equal to the projection $da_\perp$ of an elementary area $da$ of $S_2'$ onto the flat surface of the plate of the capacitor. Eventually, the integral on the right of (A.8) equals the total area $A$ of the plate. Hence, $I_{d,in}=I(t)$ and, given that $I_{in}=0$ on $S_2$ , the Ampère-Maxwell law (A.3) again yields the result (A.7).

So, everything works fine with regard to the Ampère-Maxwell law, but there is one law we have not taken into account so far; namely, the *Faraday-Henry law*! According to that law, a time-changing magnetic field is always accompanied by an electric field (or, as is often said, "induces" an electric field). So, the electric field outside the capacitor cannot be zero, as claimed previously, given that the time-dependent current $I(t)$ is expected to generate a time-dependent magnetic field. For a similar reason, the electric field inside the capacitor cannot have the static-field-like form (A.4) (there must also be a contribution from the rate of change of the magnetic field between the plates).

An exception occurs if the current $I$ that charges the capacitor is constant in time (i.e., if the capacitor is being charged at a constant rate) since in this case the magnetic field will be static everywhere. But, in the general case where $I(t)\neq$constant, the preceding discussion regarding the charging capacitor problem needs to be revised in order to take into account the entire set of Maxwell's equations; in particular, the Ampère-Maxwell law as well as the Faraday-Henry law.

## Appendix III. General form of the electric field

To justify the general expression for the electric field implied in the *Ansatz* (5) used to find solutions of Maxwell's equations inside the capacitor, we need to prove the following:

*Lemma 1.* If the magnetic field inside the capacitor is azimuthal, of the form

$$\vec{B} = B(\rho,t)\,\hat{u}_\varphi \tag{A.9}$$

then the electric field (also assumed dependent on $\rho$ and $t$) is of the form

$$\vec{E} = E(\rho,t)\,\hat{u}_z \tag{A.10}$$

*Proof.* Let

$$\vec{E} = E_\rho(\rho,t)\,\hat{u}_\rho + E_\varphi(\rho,t)\,\hat{u}_\varphi + E_z(\rho,t)\,\hat{u}_z \qquad \text{(A.11)}$$

Then (cf. Appendix I) from Gauss' law (4*a*) it follows that

$$\frac{\partial}{\partial\rho}(\rho E_\rho) = 0 \;\Rightarrow\; E_\rho \equiv \frac{\alpha(t)}{\rho} \qquad \text{(A.12)}$$

In order for the electric field to be finite at the center of the capacitor (i.e., for $\rho=0$) we must set $\alpha(t)\equiv 0$, so that $E_\rho(\rho,t)=0$. On the other hand, the $z$-component of Faraday's law (4*c*) yields

$$\frac{\partial}{\partial\rho}(\rho E_\varphi) = 0 \;\Rightarrow\; E_\varphi \equiv \frac{\beta(t)}{\rho} \qquad \text{(A.13)}$$

Again, finiteness of the electric field for $\rho=0$ dictates that $\beta(t)\equiv 0$, so that $E_\varphi(\rho,t)=0$. Eventually, only the $z$-component of the electric field is non-vanishing, in accordance with (A.10).

The solutions outside the capacitor are subject to the restriction $\rho>0$. The expression for the electric field implied in the *Ansatz* (16) is based on the following observation:

*Lemma 2.* If the magnetic field outside the capacitor is azimuthal, of the form (A.9), then the electric field (also assumed dependent on $\rho$ and $t$) is again of the form (A.10).

*Proof.* Let the electric field be of the form (A.11). Then from Gauss' law (4*a*) and from the $z$-component of Faraday's law (4*c*) we get (A.12) and (A.13), respectively. On the other hand, from the $\rho$- and $\varphi$-components of the fourth Maxwell equation (4*d*) we find that $\partial E_\rho/\partial t=0$ and $\partial E_\varphi/\partial t=0$, which means that $\alpha$ and $\beta$ are actually constants. Thus the general form of the electric field outside the capacitor should be

$$\vec{E} = \frac{\alpha}{\rho}\hat{u}_\rho + \frac{\beta}{\rho}\hat{u}_\varphi + f(\rho,t)\,\hat{u}_z \;.$$

Obviously, the function $f(\rho,t)$ is related to the time-change of the magnetic field and is expected to vanish if the current $I$ that charges the capacitor is constant. If the electric field itself is to vanish when $I=constant$, both constants $\alpha$ and $\beta$ must be zero. Eventually, the electric field outside the capacitor must be of the general form (A.10).

[1] https://www.aemjournal.org/index.php/AEM/article/view/694
[2] https://arxiv.org/abs/1711.09969
[3] https://nausivios.hna.gr/docs/2018C1.pdf